\documentclass[prl,showpacs,twocolumn]{revtex4-1}
\usepackage{psfrag}
\usepackage{amssymb,amsmath,amsthm,color}
\usepackage[dvips]{graphicx}
\usepackage{verbatim}
\usepackage[dvipsnames]{xcolor}
\usepackage{graphicx}

\usepackage{graphicx,amssymb,color}
\usepackage[dvipsnames]{xcolor}

\newcommand{\beq}{\begin{equation}}
\newcommand{\eeq}{\end{equation}}

\begin{document}
\title{Capillary-like Fluctuations of a Solid-Liquid Interface in a Non-Cohesive Granular System }

\author{Li-Hua Luu}
\email[Corresponding author: ]{luulihua@yahoo.fr}
\author{Gustavo Castillo}
\author{Nicol\'as Mujica}
\author{Rodrigo Soto}
 
\affiliation{Departamento de F\'{\i}sica, Facultad de Ciencias F\'{\i}sicas y Matem\'aticas Universidad de Chile,
Avenida Blanco Encalada 2008, Santiago, Chile}%

\date{\today}

\begin{abstract}

One of the most noticeable collective motion of non-cohesive granular matter is clustering under certain conditions. In particular, when a quasi-two-dimensional monolayer of mono-disperse non-cohesive particles is vertically vibrated, a solid-liquid-like transition occurs when the driving amplitude exceeds a critical value. Here, the physical mechanism underlying particle clustering relies on the strong interactions mediated by grain collisions, rather than on grain-grain cohesive forces.  In average, the solid cluster resembles a drop, with a striking circular shape. We experimentally investigate the coarse-grained solid-liquid interface fluctuations, which are characterized through the static and dynamic correlation functions in the Fourier space. These fluctuations turn out to be well described by the capillary wave theory, which allows us to measure the solid-liquid interface surface tension and mobility once the granular ``thermal" kinetic energy is determined. Despite the system is strongly out of equilibrium and that the granular temperature is not uniform, there is energy equipartition at the solid-liquid interface, for a relatively large range of angular wave-numbers.  Furthermore, both surface tension and mobility are consistent with a simple order of magnitude estimation considering the characteristic energy, length and time scales, which is very similar to what can be done for atomic systems. 
\end{abstract}

\maketitle

Granular matter, composed by a collection of macroscopic particles that interact  via dissipative contacts, is fundamentally out of equilibrium. When energy is injected into an athermal granular system, permanent grain rearrangements are induced and accompanied by friction and inelastic particle collisions. In particular, dense granular materials exhibit a large variety of interesting phenomena, under static conditions or in a dynamical state. Notorious examples are granular piles, avalanches, segregation, pattern formation, granular phase coexistence and jamming \cite{Jeager,Aranson,Pouliquen}. Since few decades, extensive research has been conducted on granular phase transitions. A case of interest is the solid-liquid coexistence during an avalanche, where the characterization of the two-phase interface allows the study of the granular rheology \cite{Pouliquen}. Another system of interest is vibrated granular layers that reveal Faraday waves patterns that are preceded by a solid-to-liquid transition \cite{Umbanhowar,Mujica,Mujica2}. 

Dry granular systems are usually considered to have no surface tension. However, several recent studies show that non-cohesive or very weakly cohesive granular materials develop phenomena driven by surface tension, which can be low but not zero. Some remarkable examples are the Rayleigh-Taylor-like instability in tapped powders \cite{Duran} and the interfacial instabilities in falling granular streams, in air \cite{Amarouchene} and vacuum \cite{Royer}. In other cases, granular systems are shown to behave as a zero-surface-tension liquid, as for particle sheets (analog to ``water bells'') created by a granular jet impacting a target \cite{Cheng2007} and fingering in a granular Hele-Shaw system \cite{Cheng2008}. In some cases, an hydrodynamic derivation taking the zero-surface-tension limit succeeds in describing the observations \cite{Cheng2007,Cheng2008} but in others a finite surface tension is needed \cite{Ulrich2012}. Studying the spinodal decomposition in a vibrated non-cohesive granular media, recent experiments suggest that the cluster coalescence is consistent with a curvature driven force  and numerical simulation showed stress tensor anisotropies, linked to surface tension \cite{Clewett2012}.
The point is to understand how capillary-like features can emerge out of collections of particles that are known to be almost or completely non-cohesive. 
Considering either the flow of interstitial air \cite{Duran,Amarouchene} or nano-Newton cohesion forces in the case of very low external forcing~\cite {Royer}, a low effective surface tension depending on the granular system dynamics was estimated. 

Here, we experimentally investigate a granular system composed by non-cohesive mono-disperse grains that are confined in quasi-2D geometry. This geometry allows the characterization of both individual and collective grain motion. When such system is vertically vibrated, it can present a transition from a completely fluidized state to the coexistence of a liquid state with solid (ordered) clusters \cite{prevost2004, clerc2008, Castillo}. It has been recently shown that such coexistence is triggered by a negative compressibility, as observed in a similar gas-liquid van der Waals granular transition \cite{Argentina2002,cartes}, and that density waves propagate in the system \cite{clerc2008}. In our previous work \cite{Castillo} we show that depending on the vertical height and filling density the transition can be of either first or second order type. For both cases density fluctuations do not show strong variations at the transition, but  local order varies strongly, either abruptly or continuously respectively, with a critical-like behavior in the second case.

\begin{figure}[h!]
\begin{center}
\includegraphics[width =9 cm]{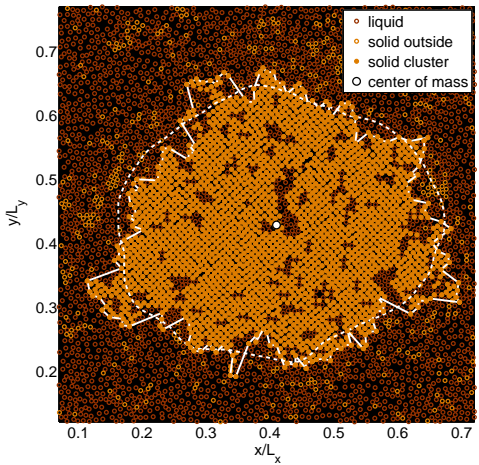} 
\caption{Map of $|Q_4^j|$ in real space. Particles are classified in the solid or liquid phase using the criterion defined in the text. The solid line with dots shows the interface detection. The dashed line corresponds to the average interface obtained from 7000 images acquired at 500 fps. The solid white circle shows the center of mass of particles in the largest solid cluster. }
\label{setup}
\end{center}
\end{figure}

In this article, we stand in the stationary regime of phase coexistence in the case where the transition is continuous. Above a critical driving amplitude a unique solid cluster is observed. As the transition is critical with an associated non-conserved order parameter, the system spontaneously develops regions of one phase inside the other.  However, in average, the granular cluster exhibits a striking circular shape, like a drop. (see Fig. \ref{setup}).  
We focus on the characterization of the liquid-solid-like interface, which we coarse-grained, becoming smooth and simply-conected.
Our approach is similar to the one proposed in \cite{Amarouchene},  in  analogy with condensed matter at molecular scale where thermal agitation induces capillary waves that deform an interface.  The coarse-graining procedure, which implementation is described below, limits the analysis of the capillary waves to large wavelengths and long time scales. Thus, in this coarse-grained description, we aim to discuss to what extent it is valid to use some the concepts effective surface tension and interface mobility.

The experimental setup is the same one reported previously \cite{Rivas,Castillo}. It consists of a container confining a layer of $N=11504$ stainless steel spherical particles in a shallow box with transverse dimensions $L_x=L_y =100d$ and height $L_z=1.94d\pm 0.02d$, where $d=1$ mm is the particle diameter (configuration 2 of our previous study \cite{Castillo}). The top and bottom glass plates confine the particles and their separation is fixed with a square frame (side walls). Particles are illuminated from below with an array of light emitting diodes and a high speed video camera enables particle detection and tracking. The surface coverage is defined by the filling fraction $\phi = N\pi d^2/4L^2=0.904$, corresponding to 31\% of volumetric filling. We submit the system to vertical sinusoidal vibrations, with displacement $z(t) = A \sin(\omega t)$, where $A$ is the vibration amplitude. Its motion is controlled with a piezoelectric accelerometer. Special care is devoted to the control of the horizontality. For this configuration, a second-order solid-liquid-like transition occurs when the dimensionless acceleration $\Gamma = A\omega^2/g$ reaches a critical value $\Gamma_c \approx 5.1$ ($f = 2\pi/\omega= 80$ Hz). In the framework of the solid cluster's characterization, the present study is done at  $\Gamma=6.30 \pm 0.03$ and $f=80$~Hz. 

Figure \ref{setup} displays typical results of the interface detection. To distinguish the liquid phase from the solid phase, which has square symmetry, we use a criterion based on $Q_4$, the 4-fold non-conserved local order parameter~\cite{Castillo}
\begin{equation}
Q_4^j = \frac{1}{N_{j}} \sum_{s = 1}^{N_{j}} e^{4i\alpha_{s}^j}.
\end{equation}
Here $N_j$ is the number of nearest neighbors of particle $j$ and $\alpha_s^j$ is the angle between the neighbor $s$ of particle $j$ and the $x$ axis.  For a particle in a square lattice, $|Q_4^j | = 1$. If $|Q_4^j| \geqslant 0.7$, which has been determined empirically \cite{suppmat}, the particle is considered in the solid phase; otherwise, it is in the liquid phase. Then, using coarse-grained polar coordinates ($\Delta \theta = 2^\circ$), we detect 180 interfacial particles in each image. The origin of the polar coordinate system is fixed at the time-averaged center of mass of the solid-like particles. The time-averaged interface exhibits a circular shape with a mean radius of $R_0/d=22.7\pm 0.4$.

We have verified that a slightly different $\Delta \theta$, say between 1.5 and 2.4, gives the same results when the spectra of radius and kinetic energy fluctuations are analyzed (see \cite{suppmat} for details). For $\Delta \theta \sim 1^\circ$, the interface detection presents errors due to the crystallographic order of the solid cluster:  occasionally the corresponding angles are close to a plane direction implying that a particle can be detected near the center of mass, making the interface very noisy. Additionally, for $\Delta\theta \gtrsim 2.4^\circ$ the detection acts as a filter for larger $m$ numbers and affects the spectrum accordingly.

In this study, we consider a curve-driven interface behavior for the steady state cluster. In condensed matter, from the classical solid to liquid phases, any interface is microscopically rough due to the competition between thermal energy and minimization of surface area \cite{Howe}. 
The grain-boundary or solid-liquid interface evolution is involved in processes such as crystallization in solution \cite{Mangin} or dendritic solidification \cite{Hoyt2003}, and controls structural and mechanical properties of many materials \cite{Yip}. 
The first interfacial parameter studied is the solid-liquid interface stiffness, $\tilde \gamma = \gamma + \gamma''$, where $\gamma$ is the surface tension and $\gamma''$ its second derivative with respect to the spatial coordinate. This is valid in the small slope approximation, as it is in our case \cite{suppmat}. In our experiment, we actually measure $\tilde \gamma$, but for simplicity we will use $\gamma$ and refer to it as surface tension. The correction $\gamma''$ is indeed usually small \cite{Hoyt2003}. The second parameter is the solid-liquid interface mobility $M$, defined by $V = M \gamma \kappa$, with $V$ the interface velocity and $\kappa$ the interface curvature.  Inspired by theoretical and numerical studies on interfacial properties of molecular systems \cite{Trautt,Hoyt}, and on an experimental study of colloidal crystals \cite{Skinner}, we attempt to obtain these physical quantities applying a capillary wave description.

In analogy to the capillary theory in condensed matter, we  assume that there is a functional, analog to the free energy, that is minimized in the stationary state and allows to obtain the dynamics close to the stationary state. This assumption, although no fully justified in non-equilibrium systems, is made for simplicity and verified \emph{a posteriori} as its predictions are consistent with the experimental results. To follow the analogy with equilibrium systems, this functional will be refered as \emph{non-equilibrium free energy}. 

First, we consider the interface contribution to the non-equilibrium free energy $E_{\gamma}$. In two dimensions, it is related to the cluster's arc length and by an effective surface tension $\gamma$ such as $E_\gamma= \gamma \int_0^{2\pi}  \sqrt{R^2+(\partial_\theta R)^2} \ d\theta$, with $R(\theta,t)$ the cluster's radius. Because of the system finite size, which implies a finite number of particles and a finite stationary radius, an additional mass term has to be added to obtain the total non-equilibrium free energy (for details see \cite{suppmat}). The radius fluctuations are defined as $\delta R(\theta,t)=R(\theta,t)-R_0$, where $R_0$ is the mean radius in time and in space. Its Fourier representation is 
\begin{equation}
\delta R(\theta,t) = \sum \limits_{m=-\infty}^{\infty}  \widetilde {\delta R}_m(t) \exp(i m \theta).
\end{equation} 
For small radial fluctuations, it is direct to show \cite{suppmat} that energy fluctuations obey
\begin{eqnarray}
\delta E &=& { \frac{\pi \lambda}{R_0}  |\widetilde{\delta R}_0|^2} +  \frac{\pi \nu}{R_0} \left( |\widetilde{\delta R}_1|^2 
+ |\widetilde{\delta R}_{-1}|^2 \right)  \\ \nonumber
&\,&+ \frac{\pi \gamma}{R_0} \sum_{|m|\geq 2} |\widetilde{\delta R}_m|^2 (m^2-1),
\label{ecn_completa_E}
\end{eqnarray}
where the first term corresponds to changes in size, the second to changes in position, which exists because the translational symmetry is not perfect in the experiment and the cluster has a tendency to remain in the center of the box, and the third term corresponds to changes of the cluster's shape. Only the last term is related to the surface tension $\gamma$ and the other two introduce new coefficients, $\nu$ and $\lambda$, that should be measured as well. From here, the static power spectrum is obtained
\begin{eqnarray}
\langle |\widetilde {\delta R}_0|^2 \rangle =  \frac{\langle K_0 \rangle R_0}{\pi \lambda}, &\quad& \langle |\widetilde {\delta R}_{\pm 1}|^2 \rangle =  \frac{\langle K_1 \rangle R_0}{\pi \nu}, \label{eq_funstat}  \\ 
\langle |\widetilde {\delta R}_{|m|\geqslant 2} |^2 \rangle &=&  \frac{\langle K_{|m|\geqslant 2} \rangle R_0}{\pi \gamma (m^2-1)}.
 \label{eq_funstat2}
\end{eqnarray}
where $\langle \,\rangle$ denotes time average. The quantity $\langle K_{m} \rangle =  \frac{1}{2} N_p m_p \langle |{{\tilde v}_m}^x|^2 + |{{\tilde v}_m}^y|^2 \rangle$ is the average horizontal kinetic energy per mode, where $N_p$ is the number of particles at the solid-liquid interface, $m_{\rm p} =4.45\pm0.01$ mg is the particle mass and ${{\tilde v}_m}^x$ and ${{\tilde v}_m}^y$ are the interface particle's velocity Fourier components \cite{suppmat}. $\langle K_{m} \rangle$ is the  equivalent of the thermal energy $k_BT/2$ per mode at equilibrium. Although related, $\langle K_{m} \rangle$ is not equal to the usually defined  granular temperature $T_g=\frac{1}{2}m_{\rm p} \langle \vec v\,^2 \rangle$, because the number of active modes is not $2N_p$.

\begin{figure}[t!]
\begin{center}
\includegraphics[width=4.25cm]{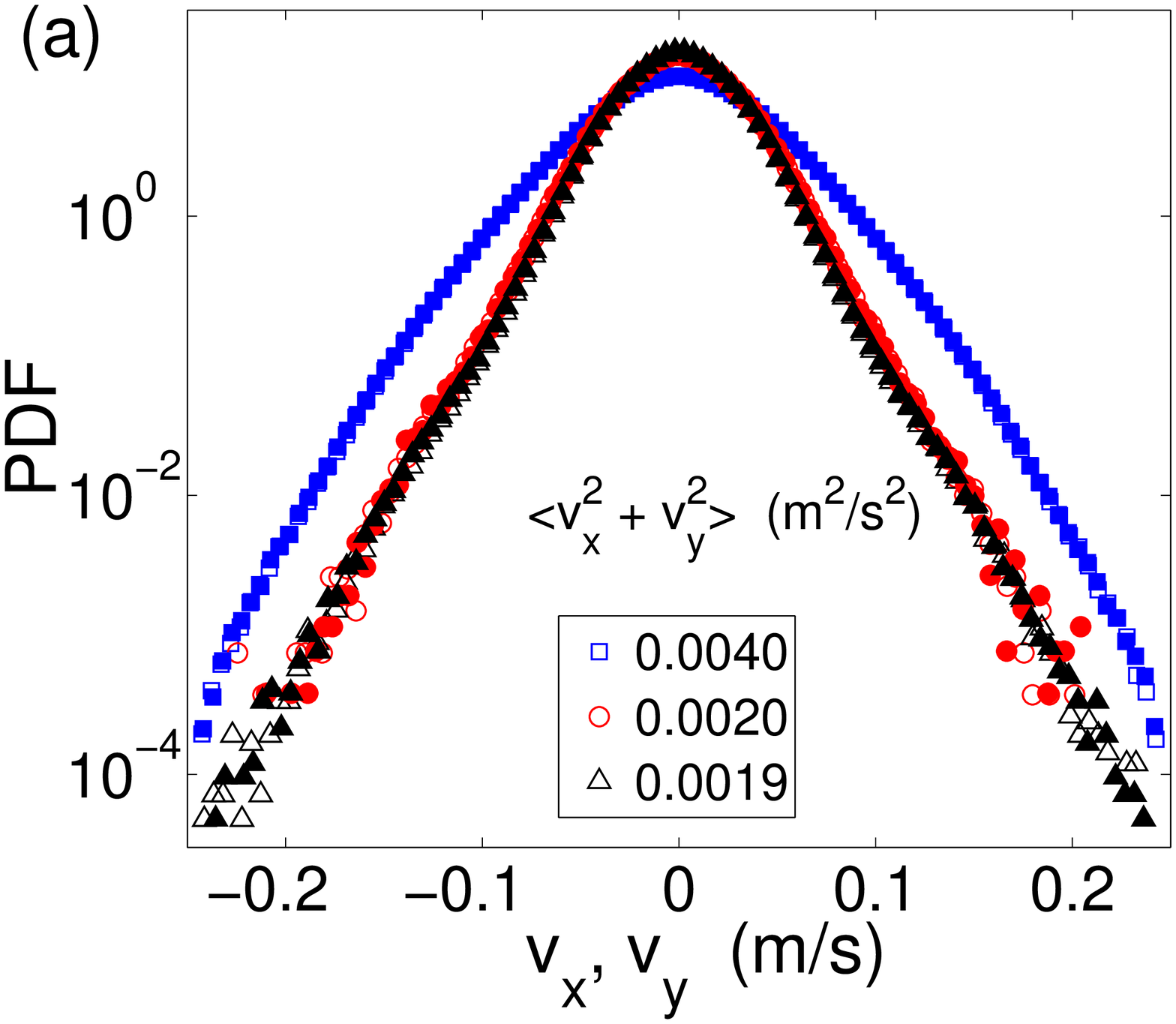}
\includegraphics[width=4.25cm]{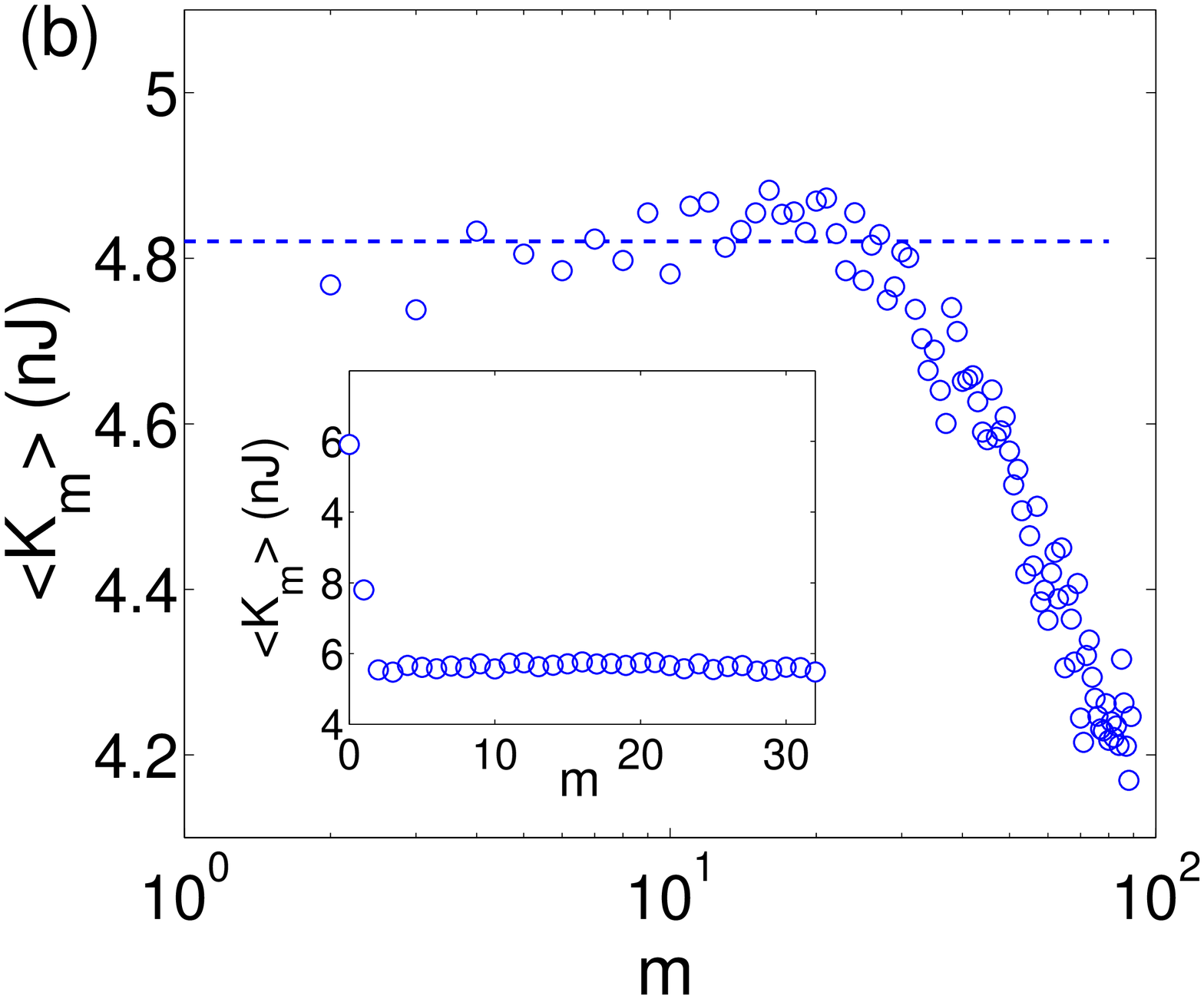}
\caption{(color online) (a)  Probability density function (PDF) of velocity fluctuations. Open and solid symbols for $x$ and $y$ components respectively. In the liquid-like domain ($\color{black} \blacksquare$), for the solid-liquid boundary particles ($\color{black} \bullet$) and in the solid-like domain ($\color{black} \blacktriangle$). (b) Average horizontal kinetic energy spectrum, $\langle K_m \rangle$ versus $m$. The dashed line shows the equipartition value $K_{\rm eq}$. The inset shows the linear plot for $m=0,...,35$.}
\label{Fig2}
\end{center}
\end{figure}

\begin{figure*}[t!]
\begin{center}
\includegraphics[width=5.9cm]{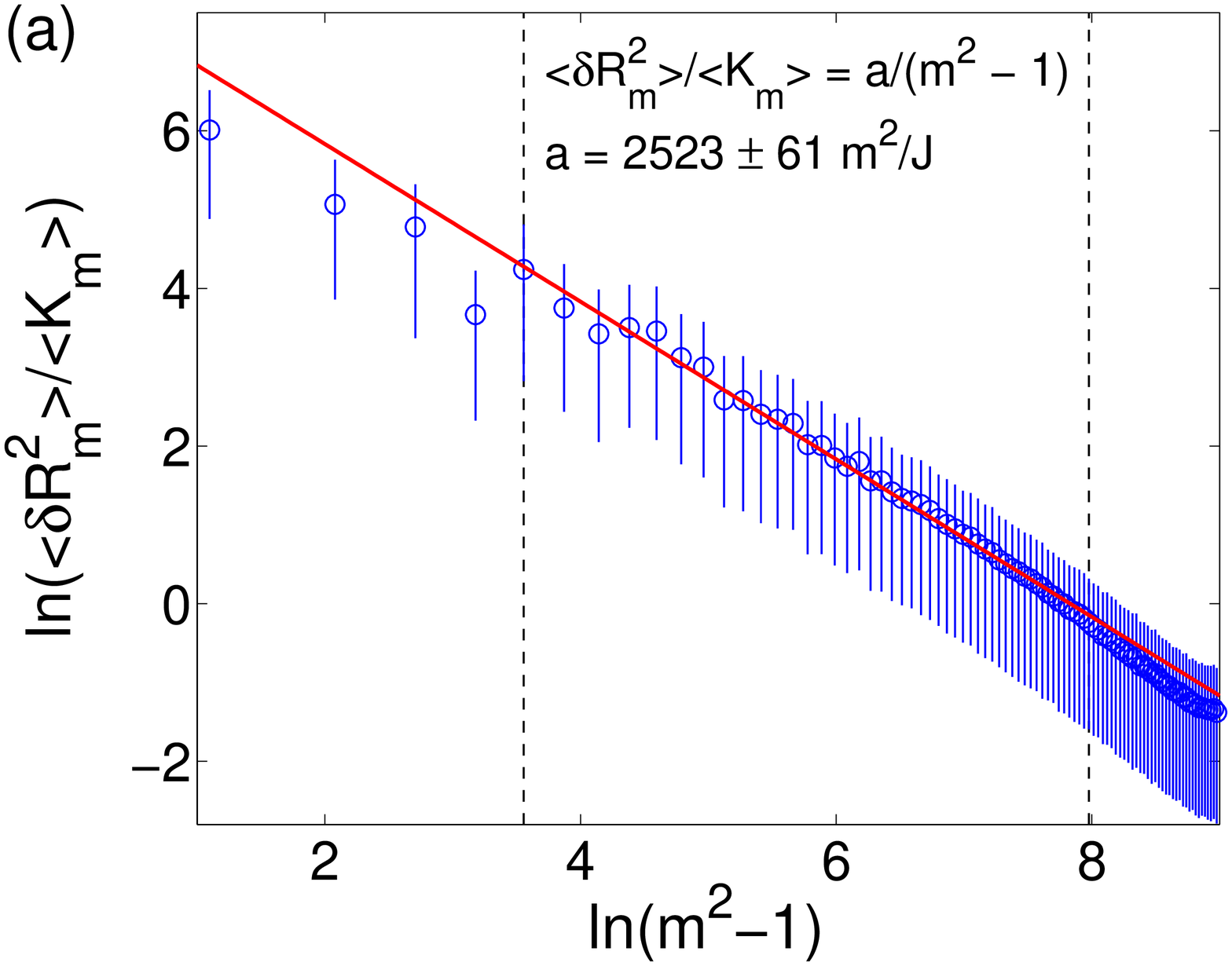}
\includegraphics[width=5.9cm]{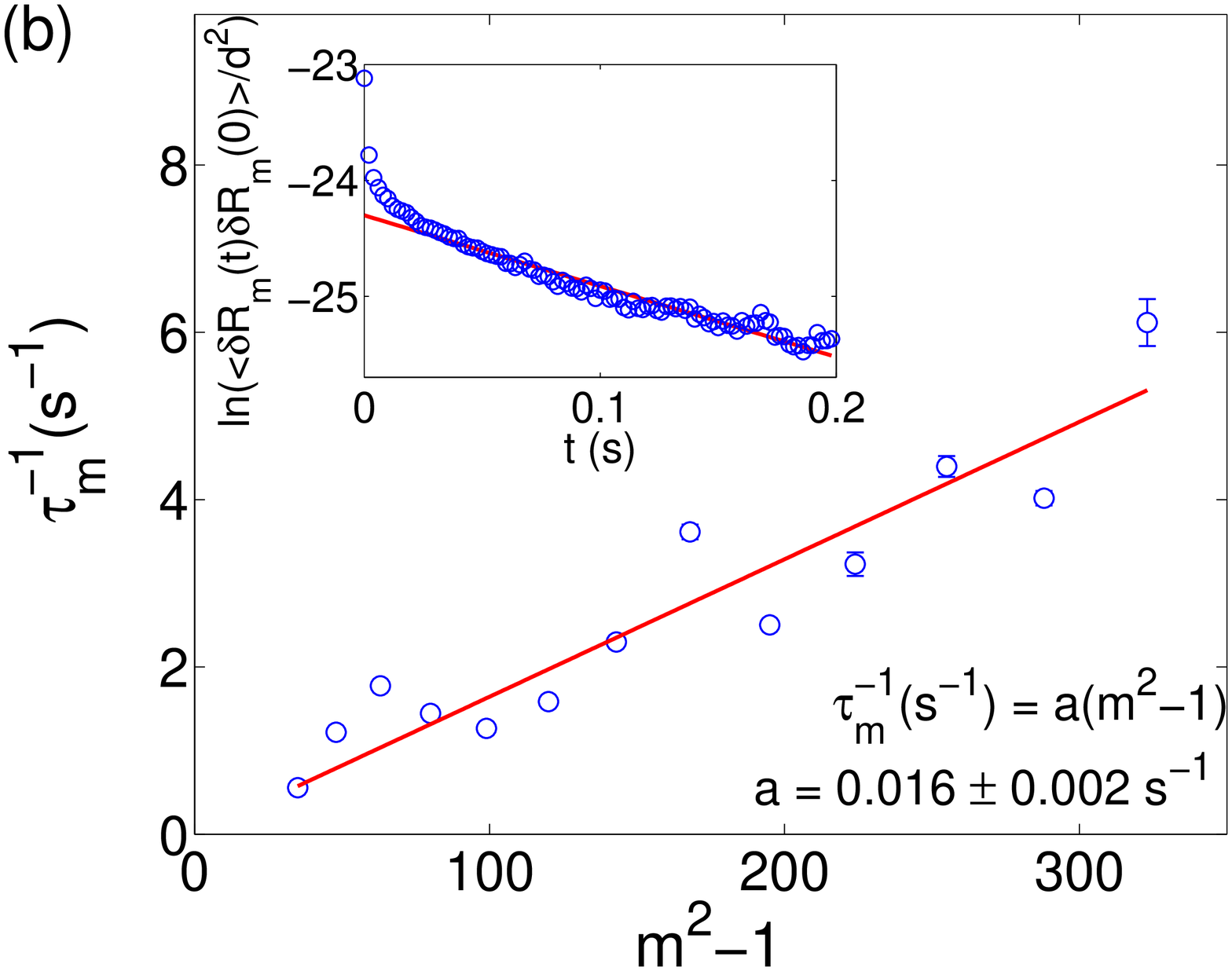}
\includegraphics[width=5.9cm]{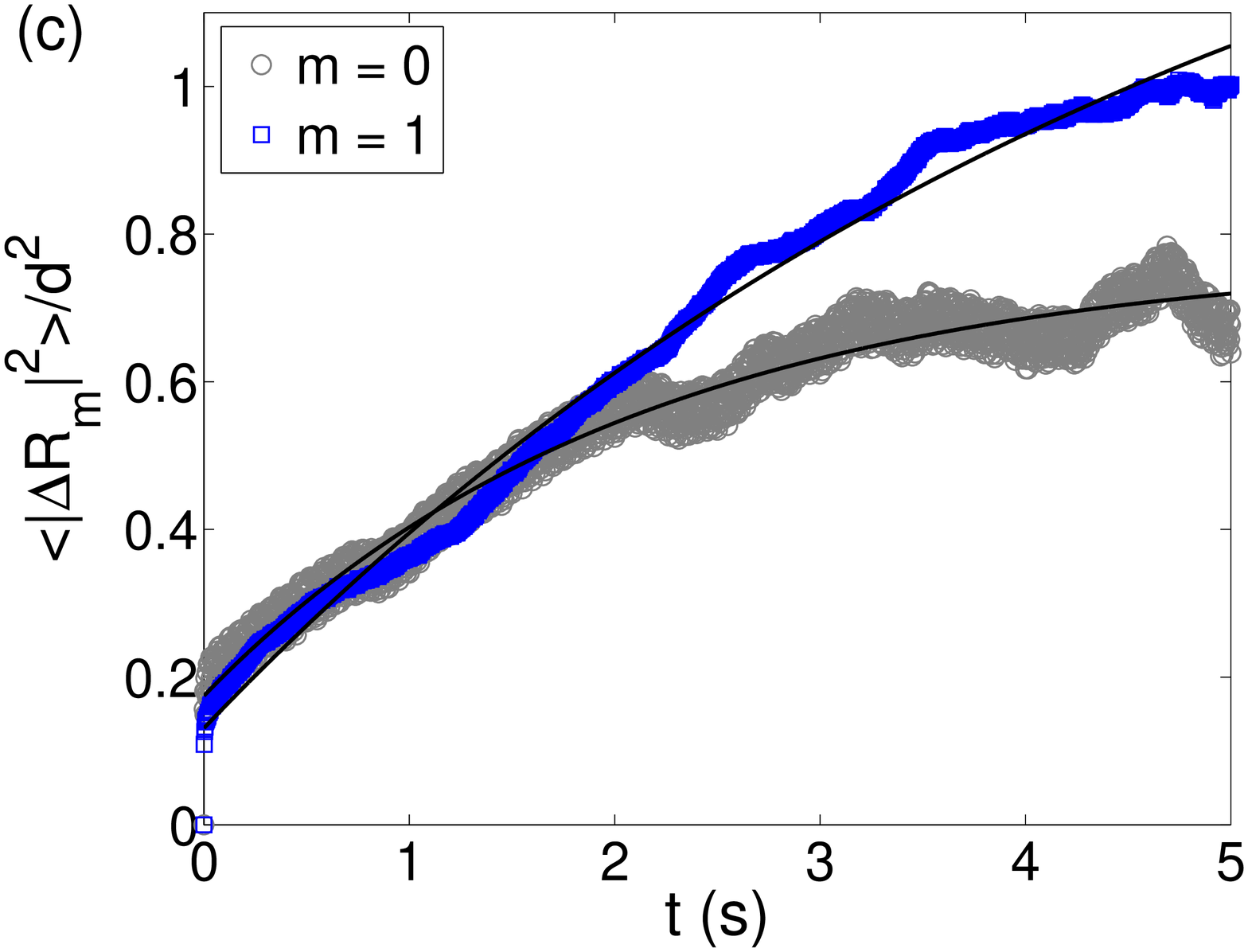}
\caption{(color online)  Static and dynamic correlation analysis. The continuous lines show the fitted functions. (a) The ratio $\langle |\widetilde{\delta R}_m|^2\rangle/\langle K_{m} \rangle$ versus $m^2-1$ in ln-ln scale. (b)  Decay time $\tau_m^{-1}$ versus $(m^2-1)$. Inset shows an example of $\langle \widetilde{\delta R}_m(t)\widetilde{\delta R}_m^*(0)\rangle$ ($m=18$). (c) Mean square displacement analysis for $m=0$ ($\circ$) and $m=1$ ($\square$). Each curve is the result of the average of two realizations. The solid lines correspond to fits $\langle | \Delta R_m|^2 \rangle = A_m [1-\exp(-t/\tau_m)] + c_m$.  The fitted parameters are $A_0/d^2 = 0.599 \pm 0.005$, $c_0/d^2 = 0.175 \pm 0.004$, $\tau_0 = 2.09 \pm 0.06$ s, $A_1/d^2 = 1.47 \pm 0.03$, $c_1/d^2 = 0.131 \pm 0.004$, $\tau_1 = 5.03 \pm 0.17$ s. }
\label{Fig3}
\end{center}
\end{figure*}

Figure \ref{Fig2}a displays the probability density function (PDF) of particle velocities in the liquid (l) and solid (s) phases, as well as for the particles at the boundary~(b). The PDFs are Gaussian for low velocities, while they present exponential tails at high velocities. Isotropy between the horizontal velocity components was verified. The granular temperature $T_g$, obtained from the variance of the velocity distributions, is not uniform. Indeed, because of the collisional and dissipative nature of particle interactions, higher particle volume fraction implies higher dissipation and lower~$T_g$. As expected, we obtain $T_g^{\rm l}>T_g^{\rm b}\gtrsim T_g^{\rm s}$, although the temperature contrast between the phases is small as compared to \cite{prevost2004} because in our case the solid cluster is soft. 

In Fig. \ref{Fig2}b we present the average horizontal kinetic energy spectrum, $\langle K_{m} \rangle$ versus $m$. Although the system is strongly out of equilibrium and that temperature is not uniform, there is energy equipartition between $m=2$ and $m \approx 30$, with an average value $K_{\rm eq}= 4.82~\pm~0.04$~nJ. For comparison, $T_g^{\rm b}~\approx~4.5$ nJ. The two lowest modes have larger energy, $\langle K_0 \rangle = 8.0\pm 4.7$~nJ and $\langle K_1 \rangle~=~5.9\pm~2.8$~nJ, and the energy components decrease for $m>30$.

Figure \ref{Fig3}a shows a very satisfactory agreement for the experimental power spectrum (\ref{eq_funstat2}) with a $1/(m^2-1)$ tendency, within a physically relevant range for the wavenumber $m = 6-54$, corresponding to wavelengths $2\pi R_0/m \approx 3d-24d$. Fitting Eqn. (\ref{eq_funstat2}) to the experimental data gives an effective surface tension $\gamma = 2.9\pm0.1$~$\mu$N in two-dimensions, and using $L_z$ as the third dimension, we get $\gamma_{\rm 3D} \equiv \gamma/L_z =1.5\pm0.1$~mN/m. Additionally, $\lambda$ and $\nu$ can be measured through the components $m=0$ and $m=1$ of $\langle |\widetilde {\delta R}_m|^2 \rangle$ and $\langle K_m \rangle$, obtaining $\lambda = 0.23 \pm 0.21$~mN, and $\nu = 0.13 \pm 0.11$~mN. 

For comparison, our measured surface tension $\gamma_{\rm 3D}$ is about 50 times smaller than pure water's surface tension, but it is much larger than the value $0.1$~$\mu$N/m estimated for a freely falling dry granular material  \cite{Royer}.  In fact, this latter work demonstrates that nano-Newton cohesive interaction forces, measured by AFM, are responsible for the surface tension. In our case, the physical mechanism is not originated by grain-grain cohesion, but from dissipative collisions between particles. Actually, the effective surface tension of our system can be estimated as $\gamma \sim T_g^{\rm b}/d\approx 4.5$ $\mu$N, which implies $\gamma_{\rm 3D} \sim T_g^{\rm b}/(d L_z) \approx 2.3$~mN/m. This result is consistent with the scaling found in a numerical study on 3D crystallization of hard spheres, where the fluid-solid surface tension is $\gamma_{\rm 3D} \sim k_BT/d^2$~\cite{Fernandez}.

From the definition of the solid-liquid interface mobility $M$, it is direct to show that each mode obeys a Langevin equation (see details in \cite{suppmat}):
\begin{equation}
\label{Ecn:Langevin_m}
\frac{\partial  \widetilde{\delta R}_{m}}{\partial t} =  - \frac{1}{\tau_m} \widetilde{\delta R}_{m} + M \eta_{m}(t),
\end{equation}
where $\tau_m=R_0^2/[M\gamma(m^2-1)]$ for $|m|\geqslant 2$, and $\tau_0 = R_0^2/(M\lambda)$ and $\tau_{\pm 1} = R_0^2/(M\nu)$ for $m=0$ and $m=\pm 1$ respectively, are the relaxation times and 
$\eta_m(t)$ are the Fourier modes of the noise term $\eta(\theta,t)$ in real space, which is assumed to be delta correlated. 
From here, the following expression for the dynamic correlation function in Fourier space is derived:
\begin{equation}
\label{eq_fundyn}
\langle \widetilde{\delta R}_m(t)\widetilde{\delta R}_m^*(0)\rangle =\langle  |\widetilde{\delta R}_m|^2\rangle e^{-t/\tau_m}.
\end{equation} 
The inset of Fig. \ref{Fig3}b displays an example of the good agreement for the predicted exponential decay. The data for $t>0.2$ s and for $m>18$ are very noisy and not considered for the analysis. 
Fig. \ref{Fig3}b shows that $\tau_m^{-1}$ does increases linearly with $(m^2-1)$. Using the value of $\gamma$, we obtain $M=2.8 \pm 0.8$ m$^3$J$^{-1}$s$^{-1}$. We remark that for an atomic system simulation \cite{Trautt}, the mobility in two-dimensions scales as $M\sim l_c^3/(k_BTt_c)$, with $k_BT$ the thermal agitation, $l_c$ and $t_c$ characteristic length and time respectively. In our case, considering $k_BT=T_g^{\rm b}$, $l_c\sim d$ and $t_c\sim d/\sqrt{\langle \vec{v}_{\rm b}^2\rangle}$, yields the same order of magnitude for the mobility $M \approx 10$~m$^3$J$^{-1}$s$^{-1}$.

The Langevin description (\ref{Ecn:Langevin_m}) can be studied considering the mean square-displacements for the $m=0,1$ modes. Indeed, from Eq. (\ref{Ecn:Langevin_m})
\begin{equation}
\langle | \Delta R_{0,1}|^2 \rangle  = \frac{2 M  \langle K_{0,1} \rangle \tau_{0,1}}{\pi R_0} \left (1-e^{-t/\tau_{0,1}} \right )\label{eq_diffusion}, 
\end{equation}
where $\Delta R_{0,1} =\widetilde{\delta R}_{{0,1}}(t) - \widetilde{\delta R}_{{0,1}}(0)$. For short times $t\ll \tau_{0,1}$, a diffusive behavior is expected for each mode, $\langle | \Delta R_{0,1}|^2 \rangle \approx 2 D_{0,1}t$, with $D_{0,1} = M\langle K_{0,1} \rangle/(\pi R_0)$, which is the analog of the fluctuation-dissipation relation used in atomic simulations for flat geometries \cite{Trautt,Hoyt}. 
Figure \ref{Fig3}c displays $\langle | \Delta R_m|^2 \rangle$ as a function of time for $m=0$ and $m=1$.  Fits are shown using $\langle | \Delta R_m|^2 \rangle = A_m [1-\exp(-t/\tau_m)] + c_m$, where $c_m$ reflects the fact that the Langevin equation does not capture the initial ballistic regime. The predicted saturation is observed for long times. From the fitted values we obtain $\lambda = 0.19\pm0.12$~mN, $\nu = 0.06\pm0.03$ mN, $M_0 = 1.3 \pm 0.9$ m$^3$J$^{-1}$s$^{-1}$, and $M_1 = 1.8\pm 1.0$ m$^3$J$^{-1}$s$^{-1}$.
The mean square-displacements are known to have poor convergent properties and the presented results were obtained using only two trajectories. Despite this numerical uncertainty, the qualitative shape of $\langle | \Delta R_m|^2 \rangle$ is reproduced and the fitted values are in the correct order of magnitude, showing the that the interface dynamics is consistent with the Langevin model.

To conclude, we have demonstrated that the liquid-solid-like interface in a quasi-2D vibrated granular system can be characterized by solid-liquid interface parameters such as surface tension and mobility.  Both quantities are consistent with a simple order of magnitude estimation considering the characteristic energy, length and time scales, which is very similar to what can be done for atomic systems.  The scaling of the effective surface tension  with the granular temperature $T_g^{\rm b}$ suggests that the particles' kinetic energy plays the role of the cohesive energy that originates capillary-like phenomenon in molecular liquids.  It would be interesting to relate the kinetic energy to a collisional pressure in each phase. 
For this purpose, the unexpected result of energy equipartition of the surface Fourier modes should be included in any theoretical approach.
 By doing so, we could handle the surface tension concept by thinking in terms of pressure difference, as defined by the hydrodynamic law of Laplace.  This would allow to couple a particle's scale study (e.g.  by accounting for collisions, cross section and contact duration) to a macroscopic description and would arise the question of particle pressure in granular media, well known for homogeneous gas fluidized beds \cite{Batchelor}, but still open for dense granular flows.

We thank F. Barra for valuable technical help and discussions. This research is supported by Fondecyt Grants No. 3120172 (L.-H.L), No. 1120211 (G.C. \& N.M.) and No. 1100100 (R.S.), and grants Anillo  ACT~127 and AIC~43.

\end{document}